\documentclass[a4paper]{article}
\usepackage{epsfig}
\usepackage{graphicx}  
\usepackage{color}     
\usepackage{comment}
\usepackage{color}
\usepackage{lscape}
\usepackage{amssymb}
\date{\today}
 \normalsize

\newcommand{\bmat}{\left(\begin{array}}
\newcommand{\emat}{\end{array}\right)}
\newcommand{\be}{\begin{equation}}
\newcommand{\ee}{\end{equation}}
\newcommand{\bea}{\begin{eqnarray}}
\newcommand{\eea}{\end{eqnarray}}

\def\gtwid{\mathrel{\raise.3ex\hbox{$>$\kern-.75em\lower1ex\hbox{$\sim$}}}}
\def\ltwid{\mathrel{\raise.3ex\hbox{$<$\kern-.75em\lower1ex\hbox{$\sim$}}}}
\def\gev{{\rm \, Ge\kern-0.125em V}}
\def\tev{{\rm \, Te\kern-0.125em V}}

\def    \be            {\begin{equation}}
\def    \ee            {\end{equation}}
\def    \bea           {\begin{eqnarray}}
\def    \eea           {\end{eqnarray}}
\def\eps{\epsilon}
\def\veps{\varepsilon}
\def\a{\alpha}
\def\b{\beta}

\def\lam{\lambda}

\def\s{\sigma}
\def\r{\rho}
\def\t{\theta}

\begin{document}
\renewcommand{\thefootnote}{\fnsymbol{footnote}}

\vspace{.3cm}

\title{\Large\bf Realization of Power Law Inflation \& Variants via Variation of the Strong Coupling Constant.
}

\author
{ \it \bf  M. AlHallak$^1$\thanks{phy.halak@hotmail.com} and  N.
Chamoun$^{2,3}$\thanks{nchamoun@th.physik.uni-bonn.de},
\\
\small$^1$ Physics Department, Damascus University, Damascus,
Syria.\\
\small$^2$ Physics Department, HIAST, P.O. Box 31983, Damascus,
Syria.\\
\small$^3$  Physikalisches Institut der Universit$\ddot{a}$t Bonn, Nu$\ss$alle 12, D-53115 Bonn, Germany.
}
\maketitle

\begin{center}
\small{\bf Abstract}\\[3mm]
\end{center}
We present a model of power law inflation generated by variation of the strong coupling constant. We then extend  the model to two varying coupling
 constants which leads to a potential consisting of a linear combination of exponential terms. Some variants of the latter may be self-consistent and can
 accommodate the experimental data
  of the Planck 2015 and other recent experiments.
\begin{minipage}[h]{14.0cm}
\end{minipage}
\vskip 0.3cm \hrule \vskip 0.5cm
\section{Introduction}

The new data from the experiment Planck 2015 \cite{Planck15} combined with the new BICEP/Keck and Planck analysis (BKP) \cite{BKP} make the contours of the combined Planck 2013+BiCEP2+WP+highL \cite{Planck2013} completely obsolete, and put quite stringent constraints on any inflationary model. In particular, regarding the tensor-to-scalar ratio $r$ and the spectral index $n_s$ parameters, they place an upper bound on the tensor amplitude which is inconsistent
with the combined 2013 contours. Moreover, combining the planck 2015 + BKP with the Baryonic Acoustic Oscillations measurements (BAO) \cite{P+BKP+BAO} provides much stronger constraints than Planck
2015 alone, and shows the constraints
$n_s = 0.968 \pm 0.006$ and  $r < 0.11$.

\begin{figure}[hbtp]
\centering
\epsfxsize=15cm
\epsfbox{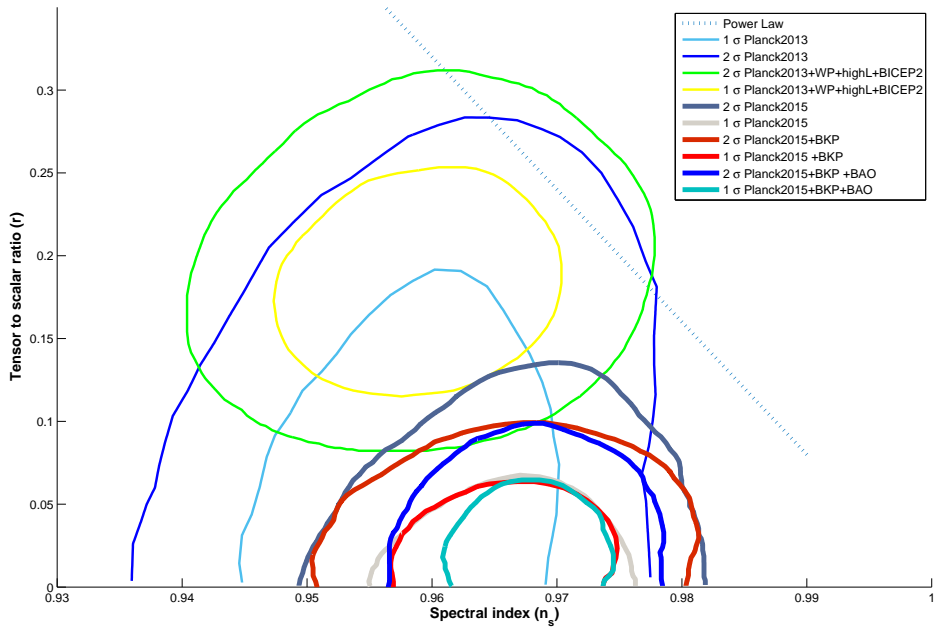}
\caption{{\footnotesize 1-$\s$ and 2-$\s$ Contour levels, for $n_s$ versus $r$, of Planck 2013 \& 2015 experiments and their Combined analyses with other experiments.}}
\label{fig1}
\end{figure}

The data of 2013 and 2015 are summarized in Fig. \ref{fig1} where the confidence contour levels, sourced from the existing ones in \cite{Planck2013} and \cite{P+BKP+BAO} respectively,  are shown for the
 Planck experiment as well as the combined analyses. Regardless of whether the BICEP2 results point to a signal of type B coming from gravitational waves originating from inflation or the signal is rather due to some dust effect \cite{sarkar}, we see that the BICEP results are consistent with Planck in a considerable region of parameter space. Consequently, we shall test our model of inflation in how it accommodates the Planck 2015 and the combined experiments.

 Inflation  \cite{Guth} is the commonly accepted theory which solves many of the Big Bang scenario problems, mainly the horizon and flatness problems. All inflationary models introduce a scalar field, the inflaton, responsible for the inflation, whose nature is still unknown. A vast array of models were studied \cite{ecyclopedia-inflation}, and they differ in the details, but most of which attribute a matter content to the inflaton. Varying speed of light (VSL), \cite{magueijo} was an alternative for solving the Big Bang scenario problems. VSL is however an integral part of ``variation of constants'' ideas \cite{vucetich-landau}. Experimental data preclude any temporal variation of the electric charge \cite{bekenstein}, the strong coupling \cite{chamoun-PLB} and the Higgs vev \cite{chamoun-JPhG} going back in time till nucleosynthesis. However, no data exist to preclude variation of constants in the inflationary era. In \cite{chamounIJMPD}, a link was suggested between variation of constants and inflation, in that it attributed the nature of the inflaton responsible for inflation to a time variation of the strong coupling constant, and the dominant contribution to energy was ascribed to quantum trace anomaly effect.

 In this letter, we re-address this link to find an exponential inflaton potential, leading to power law inflation \cite{matarrese}. Interest in power law inflation has resurfaced recently in the light of the new experiments results \cite{sahni}.  We invoke the trace anomaly just to give a rough estimate
 of the ``thermal'' gluon condensate, and contrast the predictions of the model with the experimental data.

   Being a power law inflation, the model is clearly inconsistent with the Planck 2015 \& joined experiments data because such an inflation overproduces tensor modes.
   Moreover, the model has a parameter $\ell$ with dimension of length, and in order to be even remotely consistent with data (say, with Planck 2013 as shown in Fig. \ref{fig1}),
this length scale, as we shall see, must be smaller than the Planck length $L_{\mbox{\tiny PL}}$\footnote{In this work, we omit the attribute ``reduced'' for the Planck length (mass).
The condition $\ell > L_{{\mbox{\tiny PL}}}=\sqrt{8 \pi} \sqrt{G}$, (where $G$ is the gravitational constant) is thus stricter than the version expressed in terms of the
  ``standard'' Planck length: $\ell > \sqrt{G}$.}. This is related to the well-known
``Lyth bound'', which shows that the field variation in inflation must be large
in Planck units for models  which predict a tensor/scalar ratio $r$
of order $0.1$.  Although  values for $\ell$ shorter than $L_{\mbox{\tiny PL}}$ can be envisaged in some string models \cite{bachas}, and despite the fact that the inability of
measuring a sub-Planckian length does not exclude the possibility of its existence, however the small ratio of $\ell/L_{\mbox{\tiny PL}}$ indicates that
the model needs to be supplemented by other mechanisms in order to be self-consistent and to produce inflation with the required properties.

To reconcile the power law inflation with data, one can define a modified model where coupling to gravity is non-minimal \cite{powerlawnonminimal}. Here, we do not follow
this path, but would rather assume a second gauge group with, again, a varying coupling constant. The potential becomes now a linear combination of two exponential terms.
The second varying coupling constant can be considered independent and representing a free parameter, which leads to a multi-field inflation scenario, or can be assumed
related to the variation of the strong coupling constant in such a way to suggest a form for an effective single field potential. In both cases, one can consider
variants of the power law model and study whether or not the new forms can accommodate the recent data and allow for above-Planckian length acceptable regions.

\section{The Model}
Our starting point is the ``time varying'' QCD lagrangian stated in \cite{chamoun-PLB} generalizing the work of Bekenstein \cite{bekenstein} from QED to QCD:
\bea \label{QCDlagrangian}
L&=& - \frac{1}{4} G_a^{\mu\nu} G_{\mu\nu}^a + (D_\mu\phi)^\dagger (D^\mu\phi) -V(\phi^\dagger \phi)+\frac{1}{2\ell^2} \frac{\epsilon{,_\mu} \epsilon^{,\mu}}{\epsilon^2}
\eea
where $\epsilon$ is a scalar gauge invariant and dimensionless field introduced to express the time variation of the strong coupling constant $g(x)=g_0  \epsilon(x)$, and whose dynamics is
represented by the last term of Eq. \ref{QCDlagrangian} where $\ell$ is the Bekenstein scale length and we are using the signature $(+,-,-,-)$, $a=1,\ldots,8$ runs over the gauge group generators  ($t^a$) indices, with $[t^b,t^c]=if^{abc}t^a$, the covariant derivative is defined as $D_\mu = \partial_\mu -i g_0  \epsilon(x) A_\mu $, and
the gluon tensor field is given by \bea G_{\mu\nu}^a=\frac{1}{\epsilon}\left[\partial_\mu(\epsilon A^a_\nu )-\partial_\nu(\epsilon A_\mu^a) +g_0 \epsilon^2 f^{abc} A_\mu^b A^{c}_\nu\right] \label{gluontensor}\eea

We first introduce a transformation which simplifies the formulation. By re-defining the gauge potential and the field tensor as
\bea \label{newAG} \tilde{A}^{a\mu}=\epsilon A^{a\mu} &,& \tilde{G}^{a\mu\nu}=\eps G^{a\mu\nu} \eea
we find that the explicit dependence on the field $\eps$ disappears in the definition of the new gluon field $\tilde{G}^{\mu\nu} \equiv \tilde{G}^{a\mu\nu} t^a$ in terms of the new
gauge potential $\tilde{A}^\mu=\tilde{A}^\mu_a t^a$, as
\bea
\tilde{G}^{\mu\nu} &=& \partial \tilde{A}^\nu-\partial \tilde{A}^\mu - i g_0 [\tilde{A}^
\mu,\tilde{A}^\nu]
\eea

The QCD-lagrangian can be written as:
\bea \label{newQCDlag}
L &=& - \frac{1}{4\epsilon^2} \tilde{G}^{\mu\nu}_a \tilde{G}_{\mu\nu}^a + (D_\mu\phi)^\dagger (D^\mu\phi) -V(\phi^\dagger \phi)+\frac{1}{2\ell^2} \frac{\epsilon{,_\mu} \epsilon^{,\mu}}{\epsilon^2}
\eea
in which the covariant derivative is given now by $D_\mu = \partial_\mu -i g_0 \tilde {A}_\mu $. Thus, in addition to the ``kinetic'' last term of Eq. \ref{newQCDlag}, the dependence on the $\eps$-field appears only in the gauge kinetic term divided by $\eps^2$. The equation of motion remains the same in that
 any zero variation with respect to $[\phi, A_\mu,\epsilon]$ is equivalent to a zero variation with respect to $[\phi,\tilde{A_\mu},\epsilon]$.

We will neglect the fermionic matter contribution during the inflationary era, whereas a ``thermal'' average $\langle\tilde{G}^2\rangle_T$, called henceforth ``thermal'' gluon condensate, should be taken for the gluon strength field squared in what concerns pure-gauge QCD at temperature $T$ characteristic of inflation. We get
\bea \label{QCDinfLag}
L&=&+\frac{1}{2\ell^2}\frac{\epsilon_{,\mu} \epsilon^{,\mu}}{\epsilon^2}-\frac{1}{4 \epsilon^2}\langle\tilde{G}^2\rangle_T
\eea
We shall assume the condensate $\langle\tilde{G}^2\rangle_T$ value is approximately constant during the inflationary short period in contrast to \cite{chamounIJMPD} where the constancy was assumed
to hold approximately for the condensate  $\langle{G}^2\rangle_T$.

The basic relation between the ``thermal'' gluon condensate $\langle{G}^2\rangle_T$ and the zero-temperature gluon condensate $\langle{G}^2\rangle_0$, which corresponds to a vacuum expectation value or a ground state average (estimated by QCD sum rules method \cite{shifman} to give the renormalization-independent quantity $\frac{\alpha_S}{\pi}\langle{G}^2\rangle_0 =
0.012 \pm0.004 \mbox{GeV}^4$),  can be stated as follows \cite{leutwyler}
\bea \label{basic} \langle{G}^2\rangle_T &=& \langle{G}^2\rangle_0 - \langle\rho_g-3P_g\rangle_T\eea where the temperature-dependent trace anomaly part $\langle\rho_g-3P_g\rangle_T$, with $\r_g$ ($P_g$) denotes the gluon energy density (pressure), is normalized such that it vanishes at zero temperature. For $T\succeq T_c$, where $T_c$ is QCD phase transition critical de-condensation temperature (which is of the same order of the chiral symmetry breaking $\sim 200 \mbox{MeV}$, had we introduced quarks), one can get by perturbation theory for pure gluodynamics the following \cite{boyd}:
\bea \label{boyd} \frac{\langle\r_g-3P_g\rangle_T}{T^4} &=& 4 N' b_0 \frac{N'^2-1}{288} g^4(T)\eea where for QCD with vanishing number of quark flavors we have $N'=3, b_0=\frac{11 N'}{48\pi^2}$.  This perturbative formula fits well with the lattice data available in the range up to $5 T_c$, and explains why the ``thermal'' gluon condensate becomes negative and proportional to $T^4$ as the zero-temperature contribution becomes negligible for $T>T_c$. The ideal non-interacting gas model
, which one might suspect to appear at $T> T_c$ because of ``asymptotic freedom'' in
QCD, means zero condensates at high energies, and the negative value of the condensate means that upon raising the temperature through $T_c$ , not only do the gluons coming from the interactions
in the vacuum condensate de-condense, but the further gluons which are created at the
high temperatures also take part in the de-condensation process contributing to the energy momentum trace \cite{miller}.

However, the way the gluon plasma approaches the ideal non-interacting gluon gas in the limit of very high temperatures, which restores the conformal symmetry and causes the
trace anomaly to vanish, is not clear. The lattice data are lacking for a full investigation in this region to which, presumably, inflation belongs\footnote{The technique of hard thermal
loop (HTL) perturbation theory \cite{anderson} allows also to estimate the trace anomaly. It does not agree with lattice data in the region just above $T_c$. However, the study goes up to
regions of order $10^4 T_c$ but still far short than inflation temperatures.}. We expect (look at Eq. \ref{basic}) the gluon condensate to reacquire a positive value at very high temperatures. The non-perturbative ``power corrections'', which are beyond the scope of the radiative corrections
accounted for in perturbation theory, may play an important role with effects
 in the high temperature region. In \cite{megias}, a detailed analysis of the thermal power corrections and the trace anomaly in the deconfined region was carried out. A best fit for the lattice data in the region $1.13 T_c<T<5 T_c$ was given as \bea \label{fit}
 \frac{\langle\r_g-3P_g\rangle_T}{T^4} &=& a_\Delta + b_\Delta \left(\frac{T_c}{T}\right)^2\eea with $a_\Delta=-0.04$ and $b_\Delta=3.99$. We shall extrapolate this fitting formula up to ``inflationary'' very high temperatures, and get
 \bea \label{G2T} \langle\tilde{G}^2\rangle_T &\approx& - a_\Delta \eps^2(T) T^4  \eea

In all, the gluon field thermal averaging leads to an ``effective'' potential for the $\eps$-field, and our concern is to see whether such a potential leads to inflation or not.
The positive value of the ``thermal'' gluon condensate is crucial in our analysis, otherwise we shall get an instability corresponding to a potential not bounded from below. Moreover, we need to put
the  kinetic energy term of the $\eps$-field  in its ``canonical'' form, and so we define a new field $\chi$ such that:
\bea \label{chi}
\frac{1}{2} {\chi_{,\mu}  \chi^{,\mu}} = \frac{1}{2\ell^2}\frac{\epsilon_{,\mu} \epsilon^{,\mu}}{\epsilon^2} &\Rightarrow &\eps = e^{\ell \chi}
\eea
In terms of the new field $\chi$, the lagrangian becomes
\bea \label{lagrangiancanonical}
L &=& \frac{1}{2} \partial_\mu{\chi} \partial^\mu {\chi} - \frac {1}{4} \langle\tilde{G}^2\rangle_T e^{-2\ell\chi}
\eea
We see directly here that our lagrangian can generate a power law inflation with an exponential potential
\bea \label{pot}
V(\chi) = M^4 e^{-\frac{\beta\chi}{M_{\mbox{\tiny{PL}}}}}  &,& M^4 = \frac{1}{4} \langle\tilde{G}^2\rangle_T , \beta = \frac{2\ell}{L_{\mbox{\tiny{PL}}}}
\eea where $M_{\mbox{\tiny{PL}}}$ is the Planck mass: $M_{\mbox{\tiny PL}}=\frac{1}{L_{\mbox{\tiny PL}}}\sim 10^{18} \mbox{GeV}$.

The physics behind this toy model of a varying QCD coupling leading to inflation is simple, in that we assumed a scalar field $\eps$, determining the value of
the gauge coupling, which might have assumed a value in the early universe different from its present value, and so the QCD scale is different from today. There is a contribution
to the potential of $\eps$ going, on dimensional grounds, as $\Lambda_{\mbox{\tiny QCD}}^4$, which can generate inflation provided slow roll conditions are satisfied.

\section{Analysis of the Model}
We summarize now the well known results of the power law inflation model (look, say, at \cite{ecyclopedia-inflation} and references therein).
In fact, for the power law inflation, one can find the exact solutions to the full following equations of motion ($a$ is the scale factor):
\bea \label{H}
 H^2 \equiv (\frac{\dot{a}}{a})^2= \frac{1}{3 M^2_{\mbox{\tiny PL}}} \left[\frac{\dot{\chi}^{2}}{2} + V(\chi) \right]
 &,&
 \ddot{\chi} + 3 H \dot{\chi} + \partial_\chi V =0
\eea
in the form of:
\bea \label{Hsolution}
 H(t) = H(t_{\mbox{\tiny e}}) (\frac{a(t_{\mbox{\tiny e}})}{a(t)})^{\b^2/2}
&,&
a(t)=a_{\mbox{\tiny e}} (\frac{t}{t_{\mbox{\tiny e}}})^{2/\b^2}
\eea
where the subscript ``e'' denotes the end of inflation.
We get inflation when $\ddot{a} >0$ providing $\b \leq \sqrt{2}$ which means $\frac{\ell}{L_{\mbox{\tiny PL}}} \leq 1/\sqrt{2}$, so the model does not allow for an above-Planckian length
scale $\ell$.


The e-folding number satisfies then $N- N_e = \frac{2}{\b^2} \log\frac{t}{t_e}$, whereas the time evolution of the ``inflaton'' field $\chi$ reads:
\bea \label{inflatonTimeEvolution} \chi(t) &=& \chi_e+ \frac{M_{\mbox{\tiny PL}}}{\b} \log \frac{t}{t_e}\eea We see that the field rolls down the potential from lower values to its final value $\chi_e$. The inflation duration satisfies \bea \label{infendtime} t_e^2 &=& \frac{2 (6/\b^2 - 1) M^2_{\mbox{\tiny PL}}\b^2}{M^4 e^{-\b \chi_e/M_{\mbox{\tiny PL}}}}\eea
 while the equation of state would read
\bea \label{eos} \frac{P}{\r} &=& -1 + \frac{\b^2}{3}\eea  where $P$ ($\r$) denote the pressure (energy density) of the universe.
We see that the limit $\b = 0$ (corresponding to $\ell\rightarrow 0$) leads to $P=-\r$ where the scale factor $a$ inflates exponentially.

The slow-roll parameters are defined as
\bea \label{slowroll}
\varepsilon = \frac{M^2_{\mbox{\tiny PL}}}{2} (\frac{V_\chi}{V})^2
 &,&
\eta= M^2_{\mbox{\tiny PL}} \frac{V_{\chi\chi} }{V}
\eea
where  $V_\chi=\frac{dV(\chi)}{d\chi}$  and $ V_{\chi\chi}=\frac{d^2 V(\chi)}{d\chi^2} $.

The tensor to scalar ratio and the spectral index are given by
\bea \label{rs}
r \approx 16 \varepsilon
&,&
n_s \approx 1-6 \varepsilon + 2 \eta
\eea
In our power law inflation, we get
\bea \label{rsour}
r \approx 32 (\ell/L_{\mbox{\tiny PL}})^2
&,&
n_s = 1-4 (\ell/L_{\mbox{\tiny PL}})^2
\eea
Because the slow-roll parameters are constant during inflation, then the predictions of the model do not depend on the energy scale at which the power law inflation ends.
We plot in Fig. \ref{fig1}, the line relating $n_s$ and $r$  for the choice $\b \in [0.1,0.2]$, and, as mentioned before, our power law inflationary model
can not accommodate recent data even for a sub-Planckian length scale $\ell$.

Now, the overall amplitude of the CMB anisotropies leads to an estimation of the inflation duration in that we require  \cite{ecyclopedia-inflation}
\bea \label{anisotropies} \frac{V_e^{1/4}}{M_{\mbox{\tiny PL}}} \sim 10^{-4} &\Rightarrow& M^4 e^{-\b \chi_e/M_{\mbox{\tiny PL}}} \sim 10^{-16} M^4_{\mbox{\tiny PL}} \eea
which via Eq. \ref{infendtime} and for $\b \sim 0.15$ gives an acceptable $t_e$ of order $ 10^{-33} \mbox{s}$ well earlier than the EW breaking time of $10^{-12}\mbox{s}$.

Although the observational predictions of the model cannot depend on $\chi_e$ which would be an irrelevant parameter called in just to assume an exit scenario where a tachyonic
instability is triggered, as the inflation cannot stop by slow-roll violation, however one can take by convention $\eps_e =1$ (corresponding to $\chi_e = 0$), corresponding to strong coupling constant
taking its present energy dependent value, that was reached during the inflation starting from far smaller values, and so we have $e^{-\b \chi_e/M_{\mbox{\tiny PL}}}  \sim 1$.  This helps to give a rough estimate for the reheating temperature $T_r$ at the end of inflation as $10^{-4}M_{\mbox{\tiny PL}} \sim M \sim (\langle\tilde{G}^2\rangle/4)^{1/4}$ leading to
(where we assume the ``constant'' gluon condensate still satisfies Eq. \ref{G2T} after reheating, with $\eps(T_r) \approx \eps_e = 1$):
\bea \label{Te} 3.1 \times 10^{-4} M_{\mbox{\tiny{PL}}} \sim  T_r\eea
so that $T_r \sim 10^{14} \mbox(GeV)$ .

 If we increase the value of the only free parameter $\ell$, then we see that the predictions become worse concerning the data.  It was shown in \cite{assisted} that the possibility of having several fields can support inflation even if $\b \geq \sqrt{2}$ in Eq. \ref{Hsolution}, which would mean in our case varying various gauge couplings. This motivates us to investigate in the next section variants of the  model with two varying coupling constants.

\section{Variants of the model: two varying coupling constants}

We assume that we have two groups $G_1=SU(3)_c$ and $G_2$ \footnote{One can imagine an $SO(10)$ SUSY GUT broken at $10^{16} \mbox{GeV}$ into Pati-Salam model $SU(4) \times SU(2)_L \times SU(2)_R$ which breaks at $m_{PS} < 10^{16} \mbox{GeV}$ to the LR model $SU(3)_c \times SU(2)_L \times SU(2)_R \times U(1)_{B-L}$. The final breaking from the LR into the MSSM occurs at $m_R$. Some studies \cite{Carolina} allow for a low intermediate breaking scale $m_R \sim 1 TeV$ with various high values of $m_{PS}$. If $m_{PS} \sim 10^{15} \mbox{GeV}$ then $SU(2)_R$ can play the role of $G_2$ in our case.} with varying coupling constants $g_1$ and $g_2$ expressed through two length scales $\ell_1$ and $\ell_2$ and two fields $\eps_1$ and $\eps_2$. Restricting to pure gauge, we have a generalization of Eq. \ref{QCDinfLag}:
\bea \label{2QCDinfLag}
L&=&\sum_{k=1}^{k=2} \left(\frac{1}{2\ell_k^2}\frac{\epsilon_{k,\mu} \epsilon_k^{,\mu}}{\epsilon_k^2}-\frac{1}{4 \epsilon_k^2}\langle\tilde{G_k}^2\rangle_T \right)
\eea
Again defining:
\bea \label{2chi}
\chi_k = \frac{\ln \eps_k}{\ell_k} &\Rightarrow &\eps_k = e^{\ell_k \chi_k}
\eea
we get a generalization of Eq. \ref{lagrangiancanonical}:
\bea \label{2lagrangiancanonical}
L &=& \sum_{k=1}^{k=2}\left(\frac{1}{2} \partial_\mu{\chi_k} \partial^\mu {\chi_k} - \frac {1}{4} \langle\tilde{G_k}^2\rangle_T e^{-2\ell_k\chi_k}\right)
\eea
and so the potential is expressed as:
\bea \label{2potV}
V(\chi_1,\chi_2) &=& M^4 \left(e^{-2\ell_1\chi_1} + \mu e^{-2\ell_2\chi_2}\right) \eea  where $\mu=\frac{\langle\tilde{G_2}^2\rangle_T}{\langle\tilde{G_1}^2\rangle_T}$ and
$ M^4 = \frac{1}{4} \langle\tilde{G_1}^2\rangle_T$.

\subsection{Multi-field inflation}
In order to study inflation in this double field setting, we need to introduce the corresponding slow roll parameters  as follows \cite{wands}.
\bea \label{slowroll2}
 \varepsilon_k=\frac{1}{2}\left(\frac{V_{\chi_k}}{V}\right)^2 &,& \eta_{kj}=\frac{V_{\chi_k\chi_j}}{V} \nonumber \\
 \varepsilon = \varepsilon_1 + \varepsilon_2 &,& \tan\theta = \frac{\varepsilon_2}{\varepsilon_1} \nonumber \\
 \eta_{\s\s}&=&\eta_{11}\cos^2\t+2\eta_{12}\sin\t\cos\t+\eta_{22}\sin^2\t \nonumber \\
  \eta_{s\s}&=&(\eta_{22}-\eta_{11})\sin\t\cos\t+\eta_{12}(\cos^2\t-\sin^2\t)\nonumber \\
  \eta_{ss}&=&\eta_{11}\sin^2\t-2\eta_{12}\sin\t\cos\t+\eta_{22}\cos^2\t
\eea
Then, for the slow roll regime, where all the $\varepsilon_k$ and $\eta_{kj}$'s are small, we have approximately the following for the tensor to scalar ratio and the spectral
index:
\bea \label{rs2}
r \approx 16 \varepsilon
&,&
n_s \approx 1-6 \varepsilon + 2 \eta_{\s\s}
\eea
 We shall also consider the possibility of a term representing the coupling between the two fields, and so we get the potential in the form
 \bea \label{2potW} W(\chi_1,\chi_2) &=& V(\chi_1,\chi_2) + \lam \chi_1 \chi_2
 \eea
 where $\lam$ is a positive coupling when $\mu >0$ in order to keep the potential bounded from below.
 \subsection{Single field inflation}
Alternatively, and provided one knows the actual trajectory of the compound field ${\mbox{\boldmath {$\chi$}}}=(\chi_1,\chi_2)$, one can
classically use it to get an approximative effective single field inflation. We illustrate this in two cases where we expect the path to pass through the
origin $\chi_1=\chi_2=0$ corresponding to $\eps_1=\eps_2=1$ which, conventionally, means an end to the time variation of the coupling constants (i.e. the coupling constant
 settles to its current ``energy-dependent'' law).

 \subsubsection{$\mu>0$}
Locally, one can approximate the trajectory of ${\mbox{\boldmath {$\chi$}}}$ around the origin by a straightline, and during the slow-regime
we have the equations of motion
\bea \label{slow-roll-gradient}
 H^2 \propto V
 &,&
  3 H \dot{{\mbox{\boldmath {$\chi$}}}} +  {\mbox{\boldmath {$\nabla$}}}V  \approx 0
\eea
which leads to
\bea \label{slope}
\chi_2 = \alpha \chi_1 &,& \a = \frac{\ell_2 \mu}{\ell_1}\eea
Replacing this in the Lagrangian, and defining a canonical field $\psi$ such that
\bea \frac{1}{2}\partial\chi_1^2+\frac{1}{2}\partial\chi_2^2 &=& \frac{1}{2}\partial\psi^2\eea we find
\bea \psi &=& \sqrt{1+\a^2} \chi_1\eea
Thus, the potential is given as:
\bea \label{1potW}
W(\psi) &=& M^4 \left(e^{-2\ell_1\psi/\sqrt{1+\a^2}} + \mu e^{-2\ell_2\a\psi/\sqrt{1+\a^2}}\right) \eea
Again, as the vanishing minimum is attained at infinite values of the field, an unknown parameter $\psi_e$ should interfere to mark
the appearance of new physics ending the inflation (around $\mbox{\boldmath {$0$}}$).

\subsubsection{$\mu<0$}
\begin{figure}[hbtp]
\centering
\begin{minipage}[l]{0.5\textwidth}
\epsfxsize=6cm
\centerline{\epsfbox{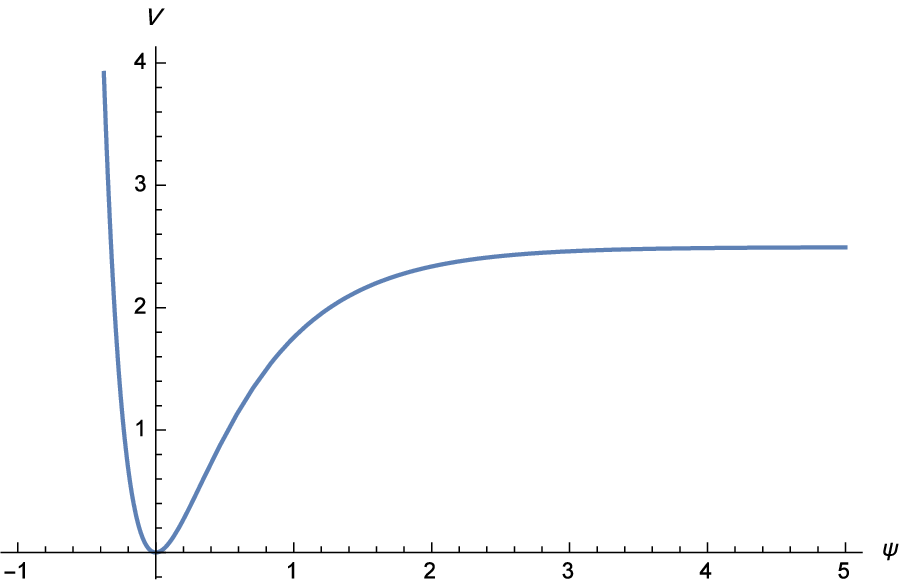}}
\end{minipage}%
\begin{minipage}[r]{0.5\textwidth}
\epsfxsize=12cm
\centerline{\epsfbox{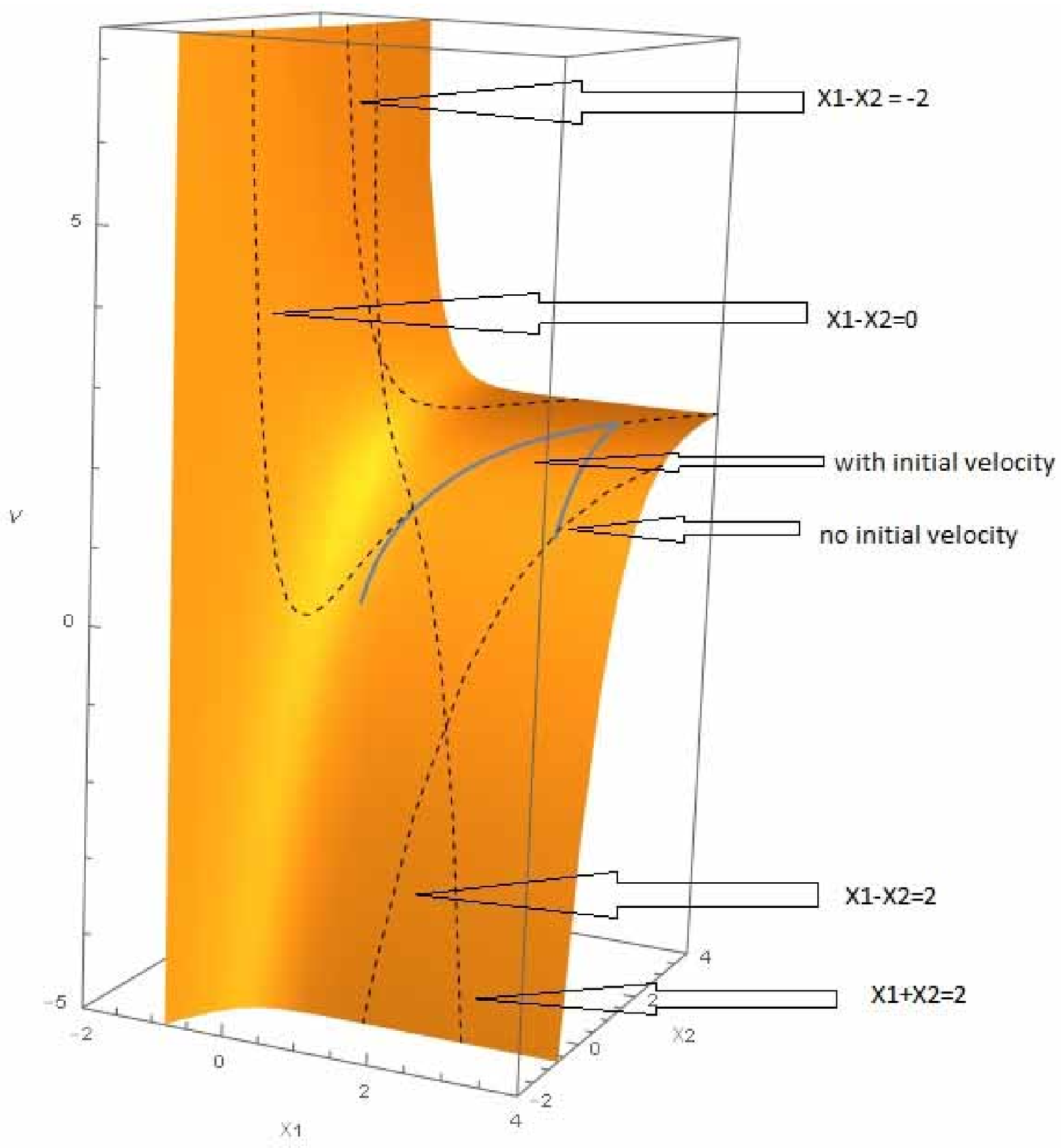}}
\end{minipage}
\vspace{0.5cm}
\caption{{\footnotesize Two Exponentials Potential. Right: 3D configuration for $V(\chi_1,\chi_2)$ with $\mu<0$. Left: $V(\psi)$  }}
\label{fig2}
\end{figure}

We do not have experimental information about the condensate of $G_2$, so we can treat $\mu$ as a free real parameter. We see that $\mu$
needs to be negative in order to get a trajectory shape with a local minimum. However, the potential with $\mu<0$ is unstable as it is unbounded
from below. We seek, in order  to simplify the study and get a single effective field inflation, a simple path which could mimic the slow roll regime of the inflation as long as possible before
the advent of the tachyonic instability, which would call for a new mechanism for an exit scenario.
Were we to assume $\eps_1=\eps_2$, then we would reobtain the power law with a new length scale $\frac{1}{l^2} = \frac{1}{l_1^2}+ \frac{1}{l_2^2}$ and a new condensate $\langle\tilde{G}^2\rangle_T = \langle\tilde{G_1}^2\rangle_T + \langle\tilde{G_2}^2\rangle_T$. Rather,
if we assume that the fields $\eps_1, \eps_2$ vary in a way that gives rise to a single canonical field $\chi_1=\chi_2=\chi$, which means:
\bea \label{epsAssumption} \eps_1^{L_{\mbox{\tiny PL}}/l_1} = \eps_2^{L_{\mbox{\tiny PL}}/l_2} &\Rightarrow& \chi_1=\chi_2=\chi,\eea
then we see in the right side of Fig. \ref{fig2} that this path ($\chi_1=\chi_2$) has a plateau followed by a minimum, resembling standard single field inflation potentials.
Defining now a new canonical field \bea \label{psi} \psi &=& \sqrt{2} \chi \eea we end up
 \bea \label{two2lagrangiancanonical}
L &=& \frac{1}{2} \partial_\mu{\psi} \partial^\mu {\psi} - M^4 e^{-\sqrt{2}\ell_1\psi} - M^4 \mu
e^{-\sqrt{2}\ell_2\psi}
\eea
and so the potential (multiplied by $-1$) is the sum of the second and third terms above. We impose now  an end to the reheating period, with the field $\psi$
at the minimum,  corresponding to  $\eps_1=\eps_2=1$,
and so the minimum should happen at $\psi=0$. This leads to $\mu = -\frac{\ell_1}{\ell_2}$ and $V(\psi=0)=1-\frac{\ell_1}{\ell_2} <0$ with $\ell_2 <\ell_1$ in order to
have a minimum. Moreover, it is true that the potential can not be defined up to an additive constant, as everything is coupled
to gravity through Einstein's equations, however the negative minimal value of the potential corresponds to an anti-de-Sitter universe. Within the multi field ``spirit''
we imagine a static third field uplifting the universe to become of Minkowski type which agrees with observations.
 Hence, we get finally the potential in the form :
 \bea \label{2Potential}
V(\psi) &=& M^4 \left[e^{-\sqrt{2}\ell_1\psi} -\frac{\ell_1}{\ell_2} e^{-\sqrt{2}\ell_2\psi} + \left(\frac{\ell_1}{\ell_2}-1\right)\right]
\eea
 This form puts the following  constraint on the condensates:
\bea \label{condensateAssumption} \frac{\langle\tilde{G_2}^2\rangle_T}{\langle\tilde{G_1}^2\rangle_T} &=& -\frac{\ell_1}{\ell_2}\eea
We do not justify here the assumptions of Eqs. \ref{epsAssumption} and \ref{condensateAssumption}, but rather take them at face value for the sake of providing a modification of the power law model making it acceptable in a way not meant to be unique. In fact, the path $\chi_1=\chi_2$ is not stable for $V(\chi_1,\chi_2)$ as the right side of Fig. \ref{fig2} shows.
However, we have simulated two trajectories (blue curves in the 3D right side of Fig. 2), starting from a point satisfying $\chi_1=\chi_2$ at the plateau
\footnote{This can be arranged, with a suitable matter content, via running of $g_1$ and $g_2$ from a common value satisfying $\eps_1=\eps_2$ at unification scale, which is higher than the inflation starting scale, to the desired initial value.}, one without an initial velocity showing clearly that the path can not
be an actual trajectory, whereas the second, with a suitably chosen initial velocity, shows that the trajectory follows the flat path long enough before falling to the instability.
One needs, however, for such trajectories to verify not only the ``potential'' slow roll conditions, but also the  conditions on the ``dynamic'' Hubble slow roll parameters, which might
require fine tuning of the initial velocity in order to assure that the path along the plateau, representing the slow roll regime, is followed long enough before the end of
inflation or before the field ${\mbox{\boldmath {$\chi$}}}$ departing from it. In all, one can thus argue that the potential $V(\psi)$ of Eq. \ref{2Potential} may reflect to a large extent the essence of the
slow roll regime, and we shall consider it our starting point with no further justification.

We can compute now the slow-roll parameters $\veps, \eta$ using Eq. \ref{slowroll} with $V$ given by Eq. \ref{2Potential} and replacing $\chi$ by  $\psi$. The canonical field $\psi$ starts at the instant $t_i$ in a large value $\psi_i$  and slowly rolls down  the potential till it reaches the end of inflation at the instant $t_e$ where the value $\psi_e$ makes $\veps \simeq 1$ or $\eta \simeq 1$ (whichever earlier):
\bea \label{psie} \psi_e &=& \max\{\psi:M_{\mbox{\tiny PL}}\frac{V^\prime}{V}=\sqrt{2} \mbox{  or  }M^2_{\mbox{\tiny PL}} |\frac{V^{\prime\prime}}{V}|=1\}\eea

During the slow-roll regime, we have the approximate equations of motion (look at Eq. \ref{H}):
\bea \label{slow-roll-H}
 H^2 \approx \frac{V(\psi)}{3 M^2_{\mbox{\tiny PL}}}
 &,&
  3 H \dot{\psi} +  V^\prime  \approx 0
\eea
and we have also $\rho \approx V$. The e-folding number is defined as
\bea \label{efoldingf} N_e &=& \int_{\psi_i}^{\psi_e} |\frac{V}{V^\prime}| d\psi\eea and we find numerically that it differs little from $N_0 = \int_{\psi_i}^{0} |\frac{V}{V^\prime}| d\psi$.

For the ``formation of structure'' problem, let us sketch now how to estimate the fractional density fluctuations using the relativistic theory of cosmological perturbations \cite{brandenberger}.
The CMB anisotropies give:
\bea
\label{fluctfinal} 10^{-5} \sim \frac{\delta \mathcal{M}}{\mathcal{M}}\mid _{t_e} &=& \frac{\delta
\mathcal{M}}{\mathcal{M}}\mid _{t_i}\frac{1}{1+\frac{p}{\rho}}\mid _{t_i}
\eea
where $\delta \mathcal{M}$ represents the mass perturbations. The
initial fluctuations are generated quantum mechanically and estimated
 by:
\bea
\label{fluctinitial} \frac{\delta \mathcal{M}}{\mathcal{M}}\mid _{t_i} &\sim& \frac{
V^\prime H}{\rho}
\eea
whence, using the continuity equation $\dot{\rho}+3(\rho+p)H=0$ and the slow roll approximation  (Eq. \ref{slow-roll-H})
\bea
\label{fluctfinal2} 10^{-5}  &\sim& \frac{1}{ M^2_{\mbox{\tiny PL}}} \frac{V' V}{\dot{\rho}} \mid _{t_i}
\eea
In order to evaluate $\dot{\rho}$, we use $\dot{\rho} \approx \dot{V} = V^\prime \dot{\psi}$ which, using again Eq. \ref{slow-roll-H}, gives us the final result \footnote{Other more refined analyses (\cite{lazarides}) lead to the same estimate multiplied by a factor of order 1}:
\bea \label{2anisotropies} 10^{-5} &\sim& \frac{\sqrt{3}}{ M^3_{\mbox{\tiny PL}}} \frac{V\sqrt{V}}{V^\prime} \mid_{t_i} \eea

This helps to get an estimate of the assumingly constant condensate:

\bea \label{condensate}
\langle\tilde{G_1}^2\rangle_T\mid_{t_e} \approx \langle\tilde{G_1}^2\rangle_T\mid_{t_i} \approx \frac{4 \times 10^{-10} M^6_{\mbox{\tiny PL}}}{3} \left( \frac{F^\prime}{F\sqrt{F}}\right)^2_{\psi=\psi_i}
&\mbox{where}& F(\psi)= \frac{V(\psi)}{M^4}
\eea

We shall not treat the reheating regime in this scenario, but just estimate $T_r$. The inflation starts with the condensate $\langle\tilde{G_1}^2\rangle_T$ getting, via
conformal anomaly, a non vanishing value at $t_i$ and remains constant henceforth (no explicit time dependence, but with an implicit one through energy dependence), and
the inflation ends at $t_e$ starting the reheating oscillatory period ending with $T_r$ \footnote{Through Eq. \ref{newQCDlag}, one can see the possibility of the decay $\chi \rightarrow \tilde{A}\tilde{A}
\tilde{A}\tilde{A}\rightarrow \Phi \Phi \Phi \Phi$ via the terms $e^{-2\ell \chi} \tilde{G}\tilde{G} \supset \chi \tilde{A}^4$ and the term $D\Phi D\Phi \supset \tilde{A}^2\Phi^2$. However,
this represents an unrenormalizable term.}.

If we assume now Eq. \ref{G2T} to remain valid at the end of reheating, then we can estimate the reheating temperature, where again $\eps(T_r) \approx 1$, as satisfying:
\bea \label{2Te}  T_r &=& \frac{10^{-2} M^{3/2}_{\mbox{\tiny PL}}}{3^{1/4}} \sqrt{\frac{F^\prime}{F\sqrt{F}}}{\mid_{\psi=\psi_i}}\eea

We can evaluate the duration of the inflation using Eq. \ref{slowroll} and get
\bea \label{Deltat} \Delta t &\approx& \int_{\psi_i}^{\psi_e} \frac{\sqrt{3}}{M_{\mbox{\tiny PL}}} \frac{\sqrt{V}}{V^\prime} d\psi\eea
We need also to check the observational bound \cite{Planck15} that during inflation and in Planckian units we have $H < 10^{-5}$, by evaluating $H$ at the end of inflation which gives
the constraint:
\bea \label{Hbound} \frac{\sqrt{F(\psi_e)} }{4 \sqrt{3}} \frac{\sqrt{\langle\tilde{G_1}^2\rangle_T}}{M^2_{\mbox{\tiny PL}}} &<& 10^{-5} \eea


\section{Discussion and conclusion}

\begin{figure}[hbtp]
\centering
\epsfxsize=15cm
\epsfbox{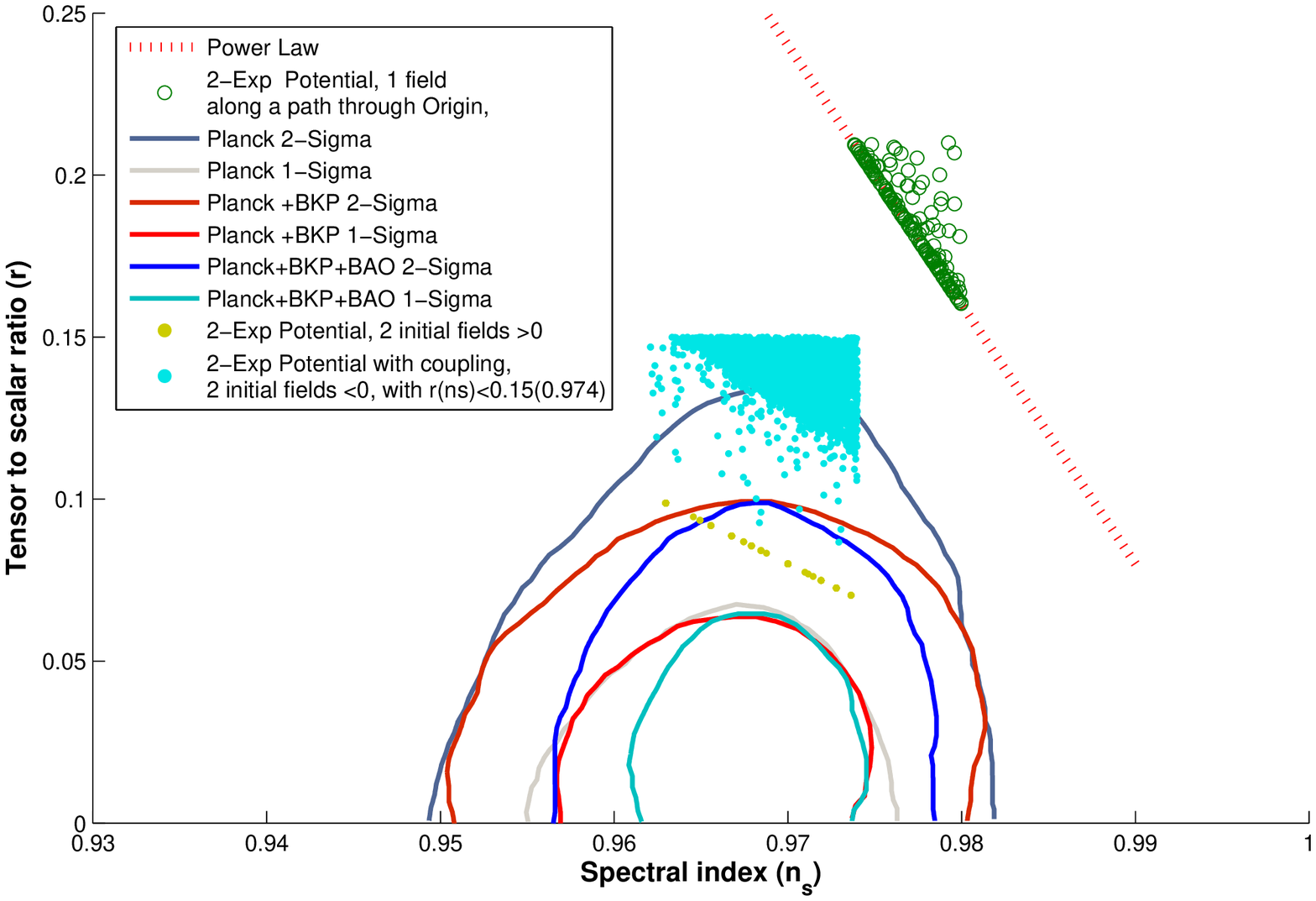}
\caption{{\footnotesize Predictions of the two-Exponentials potential, with $\mu >0$, compared to 2015 experimental data.}}
\label{fig3}
\end{figure}

Fig. \ref{fig3} represents the $n_s-r$ analysis for the three ($\mu>0$)-variants  ($W(\psi), V(\chi_1,\chi_2)$) and ($W(\chi_1,\chi_2)$) of Eqs. (\ref{1potW}, \ref{2potV}) and (\ref{2potW}) respectively.

For $W(\psi)$, the results (green colored) are lying adjacent to the power law straightline with some points further from the data. Moreover, the parameter $\ell_1$ should be sub-Planckian in order to be
nearest to the data.

For the potential $V(\chi_1,\chi_2)$, the (brown colored) points agree with data regarding ($n_s-r$). However, they are not physically acceptable, as the corresponding initial fields
are positive with very high values, and it is not plausible at all that they can be ``attracted'' to the origin, which lies at a far higher value for the potential.

For this, we introduced the potential $W(\chi_1, \chi_2)$, and although length scales are still sub-Planckian, however some points corresponding to negative-value initial fields
are acceptable physically (look at the blue colored points).

In Table \ref{tab1}, we put for each  studied pattern of $\mu>0$ the free parameters and their scanned intervals, followed by the constraints taken on the acceptable points, then the
extreme values for the acceptable points, and finally their number out of $10^7$ points randomly scattered in the scanned intervals. We opted to scan any length parameter $\ell_{1,2}$
over $[0,1] L_{\mbox{\tiny{PL}}}$ when no acceptable points exist for above-Planckian length values, whereas we chose to scan over the interval $[1,11] L_{\mbox{\tiny{PL}}}$ in the opposite case.

\begin{table}[h]
 \begin{center}
 \begin{tabular}{c|c|c|c|c|}
 \hline
 \hline
  Potential & free parameters & constraints &  Accepted boundaries & $\sharp$ accepted points  \\ \hline
  $W(\psi)$ & $\ell_1 \in [0,1]$ & $0.962<n_s<0.98$ & $[0.0708,0.0809]$ & $186$  \\
  $\propto$ & $\ell_2 \in [1,11]$ & $0<r<0.21$ & $[1.0005,10.936]$ &   \\
  $\left(e^{-2\ell_1 \psi/\sqrt{1+\a^2}}\right.$ & $\mu \in [0,1]$ &  & $[0,0.0152]$ &   \\
  $\left. +\mu e^{-2\ell_2 \a \psi/\sqrt{1+\a^2}}\right)$ & $\psi \in [-10,0]$ &  & $[-9.9326,-0.00026]$ &   \\\hline
    $V(\chi_1,\chi_2)$ & $\ell_1 \in [0,1]$ & $0.962<n_s<0.974$ & $[0.186,0.416]$ & $958$  \\
 $\propto$ & $\ell_2 \in [1,11]$ & $0<r<0.14$ & $[1.0005,10.994]$ &   \\
 $\left(e^{-2\ell_1 \chi_1}\right.$ & $\mu \in [1,10^4]$ &  & $[15.792,9.97\times 10^3]$ &   \\
 $\left. +\mu e^{-2\ell_2 \chi_2}\right)$ & $\chi_1 \in [0,10^4]$ &  & $[4.466,9.995] \times 10^3$ &   \\
  & $\chi_2 \in [0,10^6]$ &  & $[2.684, 9.959 \times 10^2] \times 10^3$ &   \\\hline
  $W(\chi_1,\chi_2)$ & $\ell_1 \in [0,1]$ & $0.962<n_s<0.974$ & $[0,0.926]$ & $63211$  \\
 $\propto$ & $\ell_2 \in [0,1]$ & $0<r<0.15$ & $[0,0.903]$ &   \\
 $\left(e^{-2\ell_1 \chi_1}\right.$ & $\mu \in [1,10]$ &  & $[2\times 10^{-4},10]$ &   \\
 $\left.+\mu e^{-2\ell_2 \chi_2}\right)$ & $\chi_1 \in [-10,0]$ &  & $[-10,-0.0006] $ &   \\
  $+\lam \chi_1 \chi_2$& $\chi_2 \in [-10,0]$ &  & $[-10, -0.0003 ]$ &   \\
  & $\lam \in [0,1]$ &  & $[0.0065, 0.999] $ &   \\\hline
   \hline
\end{tabular}
 \end{center}
 \caption{\small  Accepted points of the three variants with $\mu>0$ (out of $10^7$ scanned points). Dimensional quantities are measured in
 Planckian units ($L_{\mbox{\tiny PL}}, m_{\mbox{\tiny PL}}$).}
  \label{tab1}
 \end{table}

\begin{figure}[hbtp]
\centering
\epsfxsize=15cm
\epsfbox{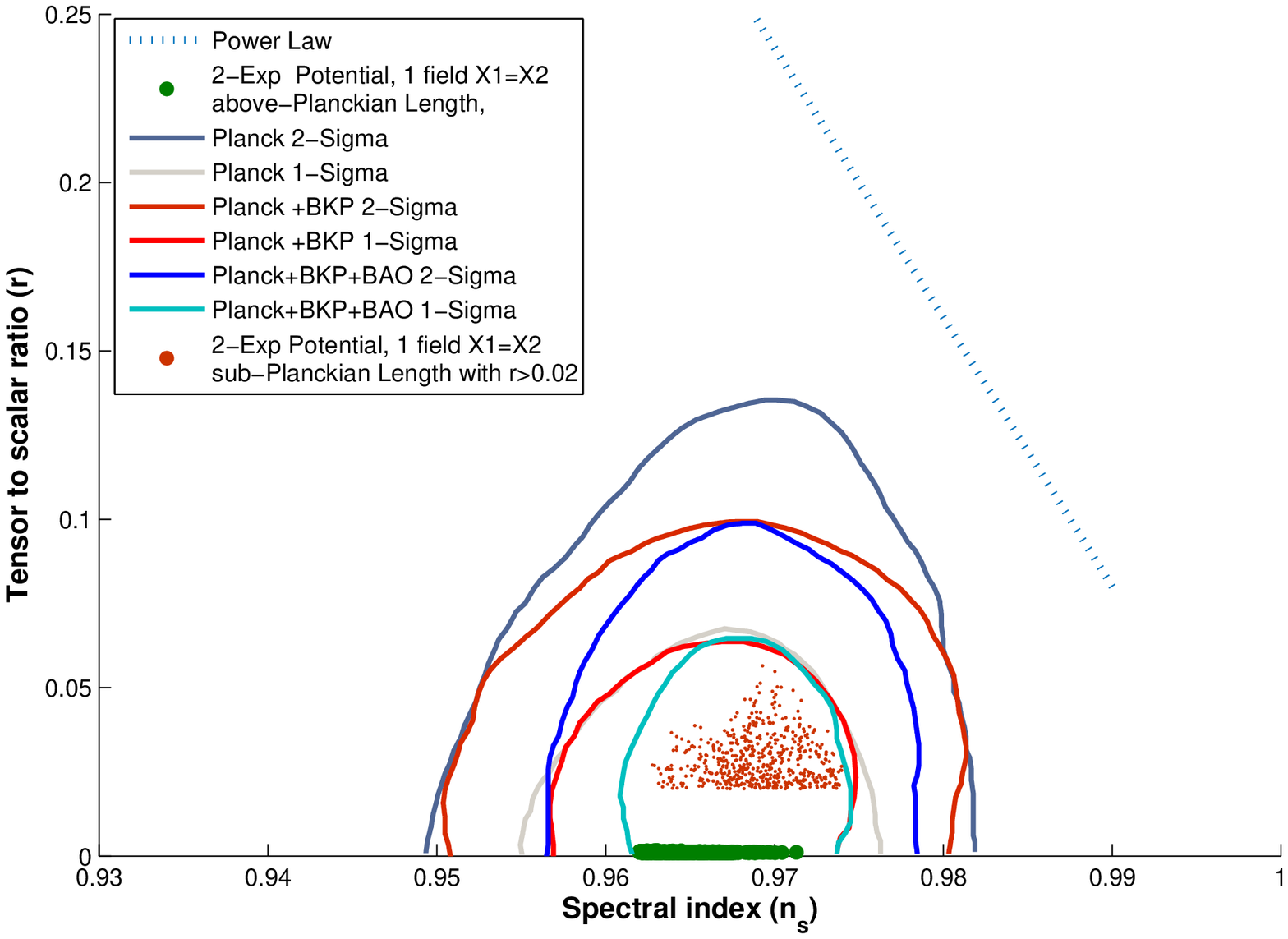}
\caption{{\footnotesize Predictions of the two-Exponentials potential, with $\mu <0$, compared to 2015 experimental data.}}
\label{fig4}
\end{figure}

\begin{figure}[hbtp]
\centering
\begin{minipage}[l]{0.5\textwidth}
\epsfxsize=8cm
\centerline{\epsfbox{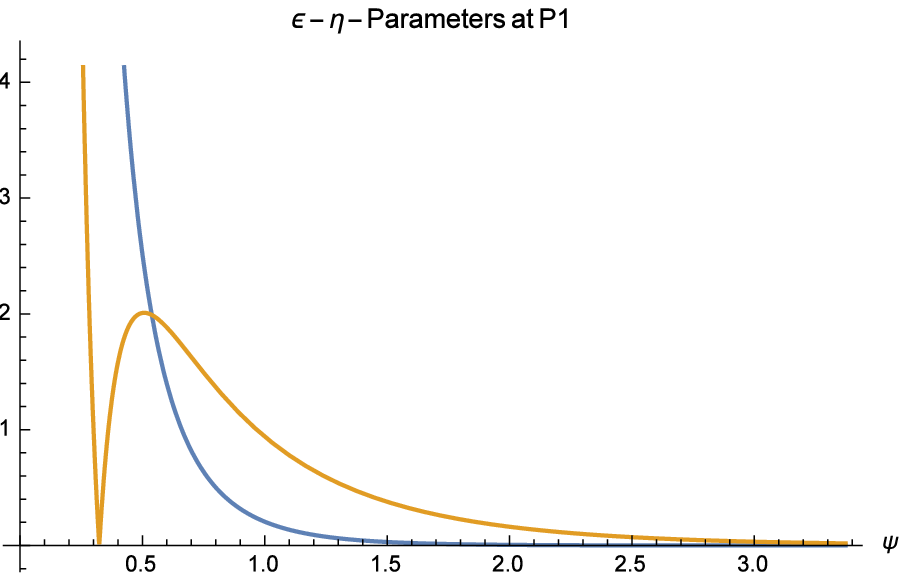}}
\end{minipage}%
\begin{minipage}[r]{0.5\textwidth}
\epsfxsize=8cm
\centerline{\epsfbox{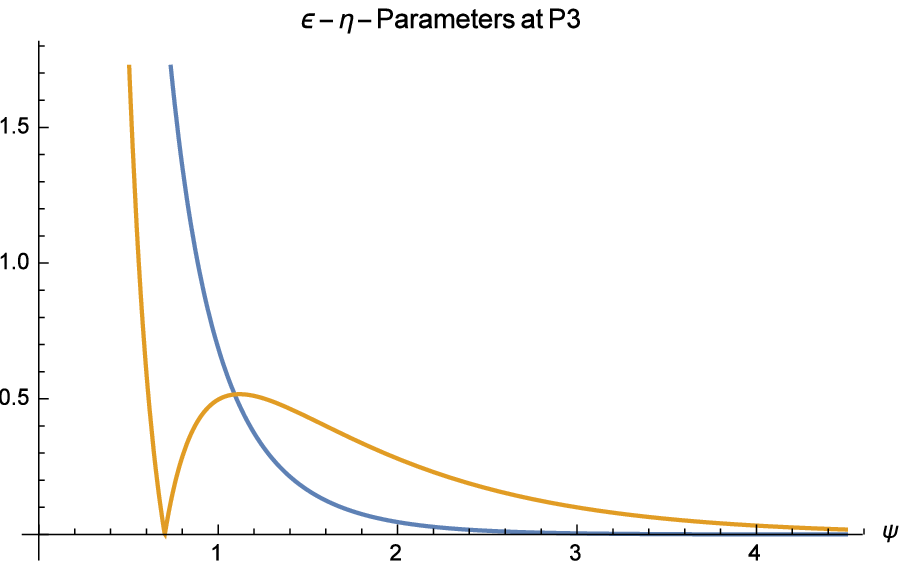}}
\end{minipage}
\vspace{0.5cm}
\caption{{\footnotesize $\eps$ (grey) and $|\eta|$ (brown)--parameters for the parameter space points P1 and P3. }}
\label{fig5}
\end{figure}

As for the pattern $V(\psi)$ of Eq. \ref{2Potential} with $\mu<0$, we carry out now a complete analysis beyond that of ($n_s-r$).
The parameter space is 3-dimensional with the free parameters $\ell_1, \ell_2$ and $\psi_i$. We take also randomly $10^7$ points in this space verifying:
\bea \label{scan} \ell_1,\ell_2 \in [1,11] L_{\mbox{\tiny PL}} &,& \psi_i \in [0,10] m_{\mbox{\tiny PL}}\eea corresponding to above-Planckian-length points,
and we forced the experimental constraints:
\bea \label{constraints} 0.962\leq n_{s_i} \leq 0.974 && 0.001 \leq r_i \leq 0.11  \nonumber \\  \ell_1 \geq l_2 && N_e \geq 50 \eea
The number of acceptable points was $1134$, and are represented by the dots colored in green in Fig. \ref{fig4}. We got an upper bound $r < 0.002$ for the points which passed the experimental constraints, and this is in line with Planck 2015 findings that for models satisfying the Lyth bound we have typically $r \leq 2 \times 10^{-5}$ for $n_s \sim 0.96$. In order to use
Eqs. (\ref{slowroll}, \ref{rs}, \ref{psie}, \ref{efoldingf}, \ref{condensate}, \ref{2Te}, \ref{Deltat}, \ref{Hbound}), we choose three benchmark points P1, P2 and P3 for this above-Planckian-length model, and summarize the findings in Table \ref{tab2}. We point out in Fig \ref{fig5} the contrast between P1 and P3 in computing $\psi_e$, in that it is $|\eta|$ $(\varepsilon)$ which reaches first the value $1$ for the point P3 (P1).

The dots colored in red in Fig. \ref{fig4} represent acceptable points which correspond to sub-Planckian lengths. For this, we scanned $10^5$ points with $\ell_1,\ell_2 \in [0,1] L_{\mbox{\tiny PL}}$, while the other constraints are kept the same, except that for visualization purposes we limited ourselves to points satisfying $0.02 \leq r_i \leq 0.11$. We got $559$ acceptable points, of which we take a representative benchmark point P4 whose results are also shown in Table \ref{tab2}.

For all the benchmark points, we see that the reheating temperature would be similar to that of the power law ($\sim 10^{14} \mbox{GeV}$), whereas the duration of the inflation would be larger, but well shorter than the EW breaking constraint ($\sim 10^{-12}$s).  Also, the observational bounds on $H_e$ are well respected in all benchmark points.

\begin{table}[h]
 \begin{center}
 \begin{tabular}{c|c|c|c|c|}
 \hline
 \hline
 & P1 & P2 & P3 & P4 \\ \hline
 $l_1/L_{\mbox{\tiny PL}}$ & $3.8280$ & $8.5857$ & $1.0091$ & $0.7249$ \\ \hline
 $l_2/L_{\mbox{\tiny PL}}$ & $1.0954$ & $2.2939$ & $1.0045$ & $0.0489$ \\ \hline
  $\psi_e/M_{\mbox{\tiny PL}}$ & $3.3725$ & $2.1375$ & $4.4985$ & $9.9499$ \\ \hline
  $\veps_i$ & $6.9\times 10^{-5}$ & $9.3 \times 10^{-6}$ & $1.1 \times 10^{-4}$ & $0.0033$ \\ \hline
  $|\eta|_i$ & $0.018$ & $0.014$ & $0.018$ & $0.0056$ \\ \hline
   $n_{s_i}$ & $0.9631$ & $0.9719$ & $0.9626$ & $0.9692$ \\ \hline
   $r_{i}$ & $0.0011$ & $1.5 \times 10^{-4}$ & $0.0019$ & $0.0522$ \\ \hline
   $\psi_{e}$ & $0.967$ & $0.848$ & $0.888$ & $1.15$ \\ \hline
   $N_{e}$ & $52$ & $70$ & $54$ & $50$ \\ \hline
    $\langle\tilde{G_1}^2\rangle_T/M^4_{\mbox{\tiny PL}}$ & $7.4 \times 10^{-15}$ & $9 \times 10^{-16}$ & $6.8 \times 10^{-12}$ & $1.3 \times 10^{-13}$ \\ \hline
 $T_r/M_{\mbox{\tiny PL}}$ & $6.5 \times 10^{-4}$ & $3.8 \times 10^{-4}$ & $36 \times 10^{-4}$ & $13 \times 10^{-4}$ \\ \hline
 $M_{\mbox{\tiny PL}} \Delta t$ & $1.6 \times 10^{16}$ & $1.6 \times 10^{17}$ & $4 \times 10^{14}$ & $6 \times 10^{14}$ \\ \hline
 $\Delta t$ (s) $\sim$ & $10^{-27}$ & $10^{-26}$ & $2 \times 10^{-29}$ & $3 \times 10^{-29}$ \\ \hline
  $10^5 H_e/M_{\mbox{\tiny PL}}$ & $0.11$ & $0.05$ & $0.23$ & $0.24$ \\ \hline
 \hline
\end{tabular}
 \end{center}
 \caption{\small  P1, P2, P3 (P4) are above(sub)-Planckian-length benchmark points, for $V(\psi)$ with $\mu <0$.}
  \label{tab2}
 \end{table}

It may seem that the model under study is very specific, and that any future data imposing a lower bound on $r>0.002$ will invalidate its above-Planckian-length version, casting doubts on it. However, one can argue that such a model is an element of an entire class of scenarios where gauge couplings at early universe might depend both on energy and explicitly
on time as well. Actually, the sub-Planckian-length benchmark point P4 can be transformed into an above-Planckian-length point by generalizing
the procedure, of going from $V(\chi_1,\chi_2)$ to $V(\psi)$, to $n$ groups with varying coupling constants under the ``strong'' assumption $\eps_k^{1/l_k} = \eps_1^{1/l_1}$. This leads to a canonical kinematic term of a field $\psi$ satisfying $\psi = \sqrt{n} \chi = \sqrt{n} \log{\eps_k}/\ell_k$, and the pure gauge potential will be given by
\bea \sum_{k=1}^{k=n} \frac{1}{4}\langle\tilde{G_k}^2\rangle_T e^{-2\ell_k \chi} &=& \frac{\langle\tilde{G_1}^2\rangle_T}{4} \sum_{k=1}^{k=n}\eta_k  e^{-\frac{2\ell_k}{\sqrt{n}} \psi}\eea
with $\eta_k = \langle\tilde{G_k}^2\rangle_T / \langle\tilde{G_1}^2\rangle_T$.
It is clear now that if the condensates satisfy: ($\eta_2 = -l_1/l_2 , \eta_k=0,k=3, \cdots, n$) then we get the same potential as in Eq. \ref{2Potential}, but now the length scales at the point P4 will be multiplied by the factor
$\sqrt{n/2}$, and for large enough $n$ one can make P4 above-Planckian-length point.

To summarize, we suggested a seemingly novel mechanism  to reinterpret inflation as a time varying coupling constant \`{a} la Bekenstein.  With only one varying coupling constant, one can realize the power law inflation which is excluded now by experiment. However, assuming more than one varying coupling constant the model opens up to several variants which can accommodate the experimental data, and may be self-consistent at the same time.


\section*{{\large \bf Acknowledgements}}
N.C. thanks E.I. Lashin for useful discussions, and acknowledges funding provided by the Alexander von Humboldt Foundation.

\end{document}